\documentclass[12pt]{article}
\usepackage{setspace,caption}
\usepackage{amsmath}
\usepackage{graphicx, amsfonts}
\usepackage{color}
\usepackage{xcolor}
\usepackage{caption}
\usepackage{amsthm}
\usepackage{bm}
\usepackage{natbib}
\usepackage{multirow}
\usepackage{ulem, cancel}

\newcommand{\XL}{\textcolor{black}}

\definecolor{darkolivegreen}{rgb}{0.33, 0.42, 0.18}

\newcommand{\widesim}[2][1.5]{
	\mathrel{\overset{#2}{\scalebox{#1}[1]{$\sim$}}}
}
\newcommand{\blind}{0}

\begin{document}
\if0\blind
{
  \title{\bf A Bayesian Nonparametric model for textural pattern heterogeneity}
  \date{}
    \author{Xiao Li\\
    Personalized Healthcare, Genentech, Inc.\\
    Michele Guindani\\
    Department of Statistics, University of California\\
    Chaan S.Ng\\
    Department of Diagnostic Radiology,\\
    The University of Texas MD Anderson Cancer Center\\
    Brian P.Hobbs\\
    Dell Medical School, The University of Texas at Austin}
  \maketitle
} \fi

\def\spacingset#1{\renewcommand{\baselinestretch}%
{#1}\small\normalsize} \spacingset{1}

\bigskip
\begin{abstract}
Cancer radiomics is an emerging discipline promising to elucidate lesion phenotypes and tumor heterogeneity through patterns of enhancement, texture, morphology, and shape. The prevailing technique for image texture analysis relies on the construction and synthesis of  Gray-Level Co-occurrence Matrices (GLCM). Practice currently reduces the structured count data of a GLCM to reductive and redundant summary statistics for which analysis requires variable selection and multiple comparisons for each application, thus limiting reproducibility. In this article, we develop a Bayesian multivariate probabilistic framework for the analysis and unsupervised clustering of a sample of GLCM objects. 
By appropriately accounting for \XL{skewness and zero-inflation} 
of the observed counts and simultaneously adjusting for existing spatial autocorrelation at nearby cells, the methodology facilitates estimation of texture pattern distributions within the GLCM lattice itself. The techniques are applied to cluster images of adrenal lesions obtained from CT scans with and without administration of contrast. We further assess whether the resultant subtypes are clinically oriented by investigating their correspondence with pathological diagnoses. Additionally, we compare performance to a class of machine-learning approaches currently used in cancer radiomics with simulation studies. 
\end{abstract}

\noindent%
\vfill

\newpage
\spacingset{1.45} 
\section{Introduction}
\label{sec:intro}
Solid masses emerge and proliferate within diverse host tissue environments making their diagnosis a challenge. With advances in scanning and high throughput computational technologies, the radiological subdiscipline of ``cancer radiomics'' has evolved in recent years \citep{yip2016applications,parekh2016radiomics}.  Facilitating non-invasive interrogation of the entire tumor volume, radiomics-derived subtypes may better describe tumor heterogeneity through patterns of enhancement, texture, morphology, and shape than biopsy based subtyping methods which inform only one or a few discrete locations. 
Thus, the interest in radiomics technologies have expanded considerably over the past decade in clinical oncology \citep{gillies2015radiomics}.

Aiming to quantitate the morphology of solid masses and elucidate its implications for prognosis, several authors have proposed machine-learning algorithms that reduce regions/volumes of interest to sets of summary statistics (or features) \citep{aerts2014corrigendum,buvat2015tumor,cook2014radiomics,lambin2012radiomics}. For example, in studying lung and head $\&$ neck (H$\&$N) cancer patients, \citet{parmar2015radiomic} employed consensus clustering to identify radiomics feature-based clusters which were subsequently validated by independent analyses. The estimated clusters were found to be significantly associated with patient survival and tumor histological profiles. \citet{chad} used consensus and hierarchical clustering to develop a radiomics signature that characterizes the local immune pathology environment of non-small cell lung cancer patients treated with definitive resection.
To differentiate asthma subgroups from 248 asthmatic patients, \citet{choi2017quantitative} applied $K$-means and hierarchical clustering algorithms with computed tomography (CT)-based structural and functional features. 
\citet{nongpiur2017anterior} employed hierarchical agglomerative clustering and a Gaussian mixture model approach on anterior segment optical coherence tomography (ASOCT) imaging data to determine if patients with primary angle-closure glaucoma could be stratified based on radiomics-derived subtypes and treatment benefit.

\subsection{Texture analysis with Grey-level Co-occurrence Matrices}
Different classes of features have been proposed and interrogated for their utility in clinical applications. Image texture analysis, the focus of this article, endeavors to describe tumour heterogeneity through patterns of gray-level spatial dependence. The predominate process for obtaining features for texture analysis involves mapping the image domain into a gray-level co-occurrence matrix (GLCM). Avoiding multivariate analysis, the GLCM object is then further summarized to produce sets of discrete, but highly redundant features for subsequent analysis via multiple regression and machine learning algorithms.

\begin{figure}
		\centering
		\includegraphics[width=0.7\textwidth,height=\textheight,keepaspectratio]{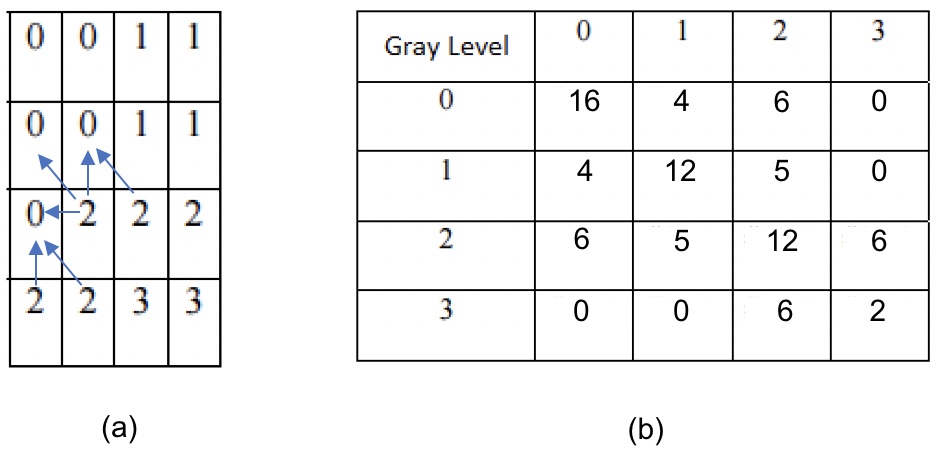}
		\caption{\small\textbf{Illustrative example of the mapping from a gray-scale image to the associated GLCM.} The original 4 by 4 gray-scale image is on the left (a), the resulting GLCM is on the right (b). The blue arrows in (a) indicate the co-occurrences of gray level 2 and gray level 0 reported in cell $(2,0)$ in (b).} 	
		\label{fig:glcm}
	\end{figure}

A GLCM is a symmetric matrix defined over an image domain with cells comprised of counts that describe the distribution of co-occurring gray-scale valued pixels (or voxels) at a given offset and all directions \citep{haralick1973textural}. More specifically, the $\left(l,h \right)^{th} $ entry of a GLCM represents the frequency with which a pixel with gray-level $l$ is present in a spatial location either horizontally, vertically or diagonally to the adjacent pixel with gray-level $h$.
{\XL{Figure~\ref{fig:glcm} provides a simple illustration of the mapping from an image in gray-scale to the associated GLCM. More specifically, figure~\ref{fig:glcm} (a) depicts a toy image with only four gray levels (0 through 3). The associated GLCM for this image is shown in figure~\ref{fig:glcm} (b), where each $(l,h)$ cell counts the number of times the gray level $l$ occurs with gray level $h$ at an adjacent neighborhood of $l$. For example, to determine the value in cell $(2,0)$ in figure~\ref{fig:glcm} (b), we count the co-occurrence of gray level 2 and gray level 0 (blue arrows) in  figure~\ref{fig:glcm} (a). }}

If we assume $K$ gray-level values (or bins), then the GLCM  contains at most $K(K+1)/2$ unique cell counts if all directions are considered. This representation facilitates image standardization, which is useful in diagnostic oncology settings wherein targeted lesions may present with irregular shapes and sizes. The complex tumor images are thus mapped to a lower-dimensional standardized lattice, consisting of $K(K+1)/2$ structured, discrete random variables, thus simplifying  model formulation and analysis. 
	
Current practice further reduces the multivariate functional structure inherent to GLCMs to sets of summary statistics (e.g., GLCM-derived textural features) for subsequent analysis via multi-variable regression and machine learning algorithms. These  reductive mappings are potentially limiting, as they may result in information loss. Moreover, many textural features are sensitive to rotations of the image and/or GLCM \citep{zhang2017study}. This lack of invariance impacts the robustness and reproducibility of feature-based approaches. Furthermore, important patterns for discerning the extent to which a malignant mass has proliferated throughout a region of interest are potentially masked with analysis based on the summary statistics.  
	
To the best of our knowledge, statistical methods able to take into account the information provided by the complete GLCM objects, instead of GLCM-derived features, are lacking in cancer radiomics. \citet{li2019spatial} have recently proposed to analyze a GLCM as a multivariate object by means of a  hierarchical Bayesian spatial model. The technique, based on Gaussian Markov random field priors, improved classification accuracy when compared with radiomics feature-based classifiers. 

As a limitation, however, \citet{li2019spatial} employed a Gaussian transformation of the observable GLCM data prior to analysis, and thus lacked a formal model for the structured count data.
{\XL {The Gaussian transformation may fail to account for the typical skewness and zero-inflation of the distribution of the GLCM counts, thereby resulting in information loss.}}
Also, the assumption of a Gaussian Markov random field intrinsically implies stationarity of the spatial process which is a restrictive assumption. Finally, \citet{li2019spatial} investigate the classification of tissue samples only in a supervised learning context, by means of binary classifiers trained by known pathology status. The objective of image-analysis is often discerning distinct patterns intrinsic to diverse tissue environments in settings wherein the specific clinical subtypes may be unknown. These limitations motivated the development of the proposed model presented herein, which extends GLCM-based textural pattern identification and analysis beyond supervised learning.

Here, we present a computationally feasible Bayesian nonparametric framework for inference of the GLCM as a multivariate {\XL {count}} object. The modeling approach captures patterns of spatial dependence intrinsic to the GLCM lattices itself, thereby avoiding further processing steps to produce reductive and redundant GLCM summary statistics. A feature-based analysis requires careful variable selection and/or multiple comparisons, and often limits the reproducibility of the resultant classifiers. By way of contrast, our proposed model for the GLCM object appropriately accounts for over and under dispersion of the observed counts, and simultaneously adjusts for existing spatial autocorrelation at nearby cells. Model specification is facilitated by a rounded kernel mixture model \citep{canale2011bayesian}, such that the observed spatial distribution of GLCM cell counts are assumed to arise from a latent continuous random vector defined on the GLCM lattice. {\XL{The rounded kernel framework  allows a flexible and robust estimation of the data generating distribution for count data\citep{Canale2017},  with respect to typical  Poisson-Gamma mixtures  \citep{karlis2005mixed, Guindani2014}}. Considering conjointly intra-lesion heterogeneity of the GLCMs objects with spatial correlation of the lattice counts, we further model the multivariate latent vectors using a spatial Dirichlet process \citep{gelfand2005bayesian}. This is achieved by defining the base distribution with a conditionally auto-regressive (CAR) Gaussian structure. Hereafter, our method is referenced as the \textit{hierarchical rounded Gaussian spatial Dirichlet process} (hierarchical rounded Gaussian SDP) model. 

Flexible for actual applications, our modeling framework facilitates pattern identification with analyses of cohorts with heterogeneous lesion sizes. 
Our unsupervised Bayesian model is applied to identify textural patterns intrinsic to adrenal lesions with CT. Data was obtained in a retrospective study of patients who underwent concordant biopsy and imaging based on a dynamic sequence of CT scans with and without administration of contrast \citep{ng3,ng1,ng2}. 
To elucidate whether textural subtypes identified exhibit pathological orientation, subtypes are evaluated for correspondence with each lesion's diagnosed pathological status. 
Additionally, simulation is used to compare performance to a class of machine-learning approaches currently promoted with cancer radiomics applications. Our Bayesian nonparametric method demonstrates robustness to capturing the underlying spatial patterns of GLCM objects, whereas commonly employed feature-based algorithmic strategies fail to identify patterns in the multivariate structure intrinsic to GLCMs.
	
The manuscript is organized as follows. Section 2 describes the motivating adrenal lesion CT imaging study. Section 3 introduces the proposed hierarchical rounded Gaussian spatial Dirichlet process model. 
Sections 4 and 5 present our case and simulation studies. 



\section{Motivating adrenal CT study}
\label{sec:meth}

Early identification and diagnosis of adrenal masses is considered pivotal clinically as it has implications for appropriate treatment selection as well as determining the prognostic status of a patient's disease stage. Proper diagnosis of abdominal lesions is also critical in the metastatic setting to determine whether a patient has experienced distant migration. Yet, characterizing benign from malignant adrenal lesions is difficult on the basis of routine CT imaging at early stages \citep{altinmakas2017diagnostic, Wanis2015}. Our study of GLCM patterns was motivated by an adrenal lesion study that comprised patients who had CT imaging available in the MD Anderson Cancer Center's radiology picture archiving and communication system (PACS) who also underwent pathological diagnoses between January 2001 and January 2010. CT images were acquired for both non-contrast (NC) and delayed post-contrast (DL) scans. CT with delayed post-contrast imaging has been well recognized in the evaluation of adrenal lesions \citep{korobkin1996delayed, boland1997adrenal, park2012washout, taffel2012adrenal}. It leverages the observation of washout characteristics of adenomas and nonadenomas after intravenous CT contrast media administration \citep{korobkin1996delayed, boland1997adrenal}.
GLCM-based clustering analyses were considered on the basis of both NC and DL scans using pixel-level data from the slice that exhibited the maximal axial cross-sectional area of the adrenal mass. Data was observed for a total of 210 adrenal lesions in 204 patients. Pathological analysis was available to establish benignity from malignancy. Of the 210 lesions, 114 were benign and 96 were malignant.
The images were reviewed using soft tissue windows (W=400; L=350), by a radiologist with more than 5 years of experience in abdominal CT imaging. For each lesion, a region of interest (ROI) was carefully drawn free-hand, with an electronic cursor and mouse, around the periphery of the adrenal lesion. As a precaution to avoid partial volume artifacts, the extreme edges of the mass were meticulously avoided. The same ROI was considered for both NC and DL scans on each subject. Then, GLCMs were constructed from the obtained ROIs as follows. Gray-level bins were constructed from the Hounsfield unit (HU) scale through evaluation of the empirical distribution of all pixels in the study, for NC scans. Pixel values below the $0.025$ and above the $0.975$ quantiles of this empirical distribution were scaled to gray-level 1 and the highest gray-level, respectively. To unfold the `contrast' effect from NC phase to delay phase, we used the same bin values from NC scans to construct delay scan GLCMs. This approach provides robustness to extreme low and high pixel outliers when projecting from the Hounsfield unit space to gray-level space. The resulting GLCMs object provide second order summary statistics of the spatial properties  of the images, and it is widely used for texture analysis in cancer radiomics.

\section{A Bayesian nonparametric model for gray-level co-occurrence matrices}
\label{sec:verify}

We start by recalling that a GLCM constructed from $K$ gray-levels is a symmetric matrix with $K(K+1)/2$ unique element counts. Counts in nearby cells are likely to be associated.  Therefore, the GLCM can also be regarded as a lattice system with a finite collection of $K(K+1)/2$ random count variables. \XL{For notational simplicity,} let $\boldsymbol{s}=\left( s_1,s_2,...,s_n\right)$ denote the GLCM locations spanning the unique cells of the symmetric structure. Correspondingly, let $Z_t(s_i)$ represent the GLCM counts at site \XL{(cell)} $s_i$ for the $t$-th subject, $t=1,2,...,T$ and $i=1,2,...,n$. To flexibly describe the integer-valued responses, we assume a rounded kernel formulation \citep{canale2011bayesian}, that is we assume there exists a latent vector $\boldsymbol{y(s)}=(y(s_1),...,y(s_n))^T$, such that the probability mass function $p$ of $\boldsymbol{Z(s)}=(Z(s_1),...,Z(s_n))^T$ can be represented by the mass assigned by the latent vector $\boldsymbol{y(s)}$ on a partition of $\mathbb{R}^n$ . More precisely, for each $n$-tuple $\bm{k}=\{k_1, \ldots, k_n\}\in \mathbb{N}^n$, we can define a disjoint partition of the latent space $\mathbb{R}^n$,  with partition sets defined as \XL{$A_{\bm{k}}= \{ \boldsymbol{x} \in \mathbb{R}^n :a_{k_1}\le x_1 < a_{k_1+1},...,a_{k_n}\le x_n < a_{k_n+1} \}$}, for a sequence of cut-points $a_1, a_2, \ldots,$ such that 
	\begin{eqnarray*}
		p(Z(s_1)=k_1,...,Z(s_n)=k_n)=\int_{A_k}\, f(\boldsymbol{y(s)})\, d(\boldsymbol{y(s)}),
	\end{eqnarray*}	
where $f(\cdot):\mathbb{R}^n \to \mathbb{R}$ succinctly denotes the $n$-dimensional density distribution of the latent vector $y(\cdot)$.  Hereafter, we choose 
the cut-point parameters $a_{k_i}$ as $a_0=-\infty$ and $a_{k_i}=k_i-1$ for $k_i \in \{1,2,...\}$. 
	
An essential feature of cancer radiomics data is that  patients undergo diagnostic scanning with lesions of varying sizes. The total number of pixels within a targeted region of interest (ROI) determines the upper bound for the count space of each GLCM cell and thereby needs to be accounted for in the analyses intended to capture the co-occurrence patterns of GLCM objects, and to compare patterns across subjects. To adjust for the image size effect, we can specify the distribution of the latent GLCM counts as a mixed-effects model, as follows
	\begin{eqnarray*}
		y_t(s_i)= \boldsymbol{x_t}\boldsymbol{\beta} + \gamma_t\,\theta_t(s_i) + \varepsilon_t (s_i)
	\end{eqnarray*}
where $\boldsymbol{\beta}=(\beta_1,...,\beta_p)$ is a $p \times 1$ vector of coefficients capturing the effect of available subject-specific covariates $\boldsymbol{x_t}$, $\bm{\gamma}_t$ is a 
scaling factor, more precisely the sum of counts for subject $t$  normalized over the whole cohort, and $\boldsymbol{\theta_t}=(\theta_t(s_1),...,\theta_t(s_n))$ is a spatial random effect, which captures the spatial dependence of the GLCM measurements. 
The term $\varepsilon_t(s_i)$ denotes an i.i.d. error term, which is assumed as normally distributed with mean 0 and variance $\tau^2$. As a result, the latent variable is ensured to be a continuous, normally distributed, random variable conditional on the realization of the process $\bm{\theta}_t$. 
	
	\begin{figure}
		\centering
		\includegraphics[width=1.3\textwidth,height=0.9\textheight,keepaspectratio]{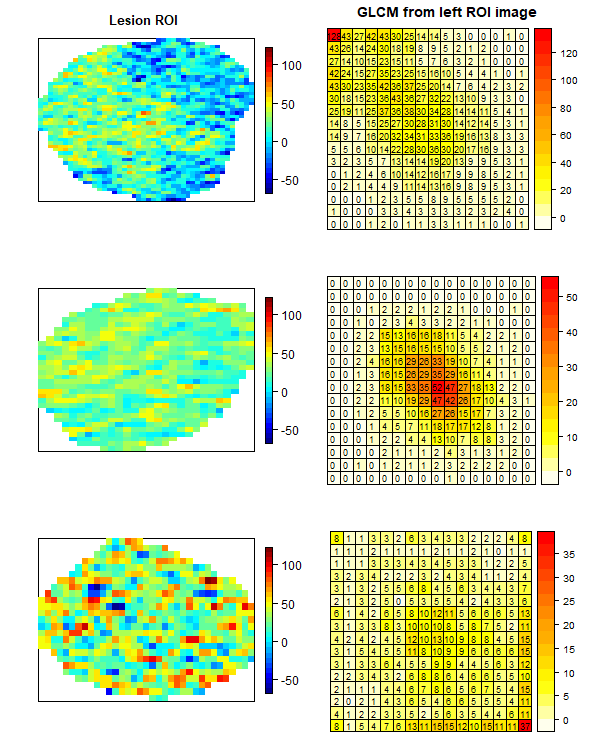}
		\caption{\small\textbf{Illustrative patterns of GLCM in adrenal lesions from non-contrast CT imaging} for 3 representative subjects, pixel-level ROIs on the left, GLCMs on the right. } 	
		\label{fig:example}
	\end{figure}
	
\subsection{Modeling the heterogeneity of the adrenal lesions} Figure \ref{fig:example} depicts enhancement patterns on the Hounsfield unit domain of ROIs comprising adrenal lesions as well as their corresponding gray-level co-occurrence matrices for three representative subjects. The top image, reflects a high degree of co-occurrence at higher gray levels, which exploratory analyses suggest to be representative of cases of malignancies within our case study. Conversely, the bottom image tends to be indicative of benign lesions.  The central image, for which high co-occurrences were observed at middle gray-levels, depicts an intermediate case.
Importantly, the GLCM cells exhibit high correlation with their adjacent neighbors, as depicted by the fairly smooth changes observed when transversing away from the peak cell counts. This suggests that lesion heterogeneity may be differentiable by spatial pattern in the GLCM object by leveraging this dependence structure with an appropriate multivariate model. Here, we propose to describe the observed heterogeneity of the GLCM matrices by modeling the random effects $\boldsymbol{\theta_t}$ with a spatial Dirichlet process \citep{gelfand2005bayesian}. In Bayesian modeling,  nonparametric priors have been often employed for unsupervised learning, to achieve improved inference on subject-specific parameters by borrowing information within population subgroups. The spatial Dirichlet process assumes that each spatial random effect $\boldsymbol{\theta_t}$ is a sample from an a.s. discrete probability distribution $G$,
	\begin{align*}
	\boldsymbol{\theta_t} \mid G \widesim{iid} G , \qquad  t=1,2,..,T.
	\end{align*}
	This is a realization of a Dirichlet Process,
	$$
	G \sim DP(\upsilon G_0), 
	$$
where $G_0$ denotes a base (or \textit{centering}) parametric spatial model, such that $E(G(A))=G_0(A)$ for any measurable set $A \subset \mathcal{R}^n$, and $\nu >0$ is a \textit{precision} parameter, characterizing the prior uncertainty of $G$ around the parametric model $G_0(\cdot)$. Since $G$ 
is almost surely discrete, there is a positive probability that some of the vectors $\boldsymbol{\theta_t}$'s take the same value for some of the subjects, thus inducing a probability-based clustering of the GLCM matrices in homogeneous subgroups.  
	
Textural patterns in the GLCM object can be identified through  model specifications that acknowledge  spatial dependencies among adjacent cell counts. In particular, here we specify the base parametric model $G_0$ as a $n$-variate normal distribution by assuming a conditionally auto-regressive (CAR) structure \citep{besag1974spatial, banerjee2014hierarchical}
	\begin{eqnarray*}
		G_0(\cdot \mid \sigma^2,\rho)=MVN(\cdot \mid \boldsymbol{0},(\boldsymbol{D}-\rho\boldsymbol{W})^{-1}\sigma^2),
	\end{eqnarray*}
where $\boldsymbol{W}$ is an adjacency matrix with $W_{kd}$ indicating whether lattice elements $k$ and $d$ are adjacent neighbors; $\boldsymbol{D}$ is a diagonal matrix with $k$th entry denoting the number of neighbors for lattice $k$; $\rho$ is a smoothing parameter controlling the degree of spatial dependence and $\sigma^2$ is a variance parameter \citep{banerjee2014hierarchical}. The covariance structure implied by the spatial Dirichlet process construction is of particular relevance. As a matter of fact, integrating out the unknown $G$, the marginal covariance matrix of the latent vector $\bm{y}(s)$ can be computed as
	\begin{eqnarray*}
		cov(\boldsymbol{y(s)}) &=& \gamma^2\sigma^2(\boldsymbol{D}-\rho\boldsymbol{W})^{-1} + \tau^2\boldsymbol{I},
	\end{eqnarray*}
i.e. constant over the entire spatial domain. However, any realization $G$ of the spatial Dirichlet process is a.s. discrete, i.e. $G=\sum_{l=1}^{\infty}\, p_l\, \delta_{\bm{\theta^*_l(s)}}$, where $\delta_{\theta}$ denotes a point mass at $\theta$, $w_l$ is a sequence of weights, and $\bm{\theta_l^*(s)}$ are sequences of unique realizations from the base distribution $G_0$ (independently of the weights). Then, for any given $G$, the conditional covariance matrix of $\bm{y}(s)$ has elements 
	\begin{eqnarray*}   
		cov(y(s_i),y(s_j)|G) &=& \gamma^2[\sum_{l=1}^{\infty}p_l\theta^*_l(s_i)\theta^*_l(s_j)-(\sum_{l=1}^{\infty}p_l\theta^*_l(s_i))(\sum_{l=1}^{\infty}p_l\theta^*_l(s_j))].
	\end{eqnarray*}
Therefore, the covariance changes over the spatial domain, and in general it is not constant in any two locations. Thus, the spatial Dirichlet process formulation provides a flexible model to describe the spatial dependence observed in the GLCMs, including ``non-stationary'' patterns. Furthermore, the effect of the scaling factor $\gamma$ on the covariance matrix reflects the general idea that regions of smaller size are characterized by smaller variance of the observed gray-scales over the spatial domain.\\
	
The hierarchical model is fully specified with the prior distributional assumptions for the remaining  parameters. More specifically, for the regression coefficients, we can assume a multivariate Gaussian prior with mean $\bm{\beta_0}$ and covariance matrix $\bm{\Sigma_{\beta}}$. To promote maximal data learning, it is common to set $\bm{\beta_0}$  as $\bm{0}$ and $\bm{\Sigma_{\beta}}$ as a diagonal matrix $\sigma^2_{\beta}\, \bm{I}$ with large $\sigma^2_{\beta}$, to allow a vaguely non-informative specification spanning a large region of the parameter space. 
The parameters $\tau^2$ and $\sigma^2$ are assigned, respectively, (conditionally) conjugate inverse-gamma priors, $IG(\tau^2\mid a_\tau,b_\tau)$ and $IG(\sigma^2\mid a_\sigma,b_\sigma)$ \XL{(with means $b_\tau/(a_\tau-1)$ and  $b_\sigma/(a_\sigma-1)$, respectively)}.  The hyperparameters $a_\tau$, $b_\tau$, $a_\sigma$, $b_\sigma$ are typically fixed to represent vague prior information.
A $Gamma(a_\upsilon,b_\upsilon)$ prior \XL{(with mean $a_\upsilon/ b_\upsilon$)} is imposed for $\upsilon$ as the corresponding mixture of gamma full conditional facilitates MCMC computation \citep{escobar1995bayesian}. 
Lastly, we follow prevailing literature and consider a  $Uniform(0,1)$ distribution for the $\rho$ parameter in the CAR model \citep{lee2013carbayes}.

Figure \ref{fig:model} provides a graphical representation of the proposed hierarchical model for the GLCM counts. Conditional on the latent $\boldsymbol{y_t}$'s, the sampling distribution can be expressed as  
	\begin{eqnarray*}
		\noindent P(  \boldsymbol{Z}|\boldsymbol{y})   &\propto & 
		\prod_{t=1}^{T} \prod_{i=1}^{n} I\left\lbrace a_{z_t(s_i)} \le y_t(s_i) <  a_{z_t(s_i)+1} \right\rbrace,\\
	\end{eqnarray*}
where $I(\cdot)$ denotes the indicator function, for fixed cut-points $a_0$, $a_1$, \ldots $a_{\max_t \,\{ z_t\}}$. Then, the hierarchical model on the latent vector $\boldsymbol{y_t}$ 
can be summarized as follows:
	\begin{align*}
	\noindent \boldsymbol{y_t} \mid \boldsymbol{\beta},\boldsymbol{\theta_t},\tau^2 &\sim MVN\left(\boldsymbol{y_t}\mid \boldsymbol{x_t}\boldsymbol{\beta}+\gamma_t\boldsymbol{\theta_t},\tau^{2}\boldsymbol{I}\right), \qquad  t=1,2,..,T \\	
	\boldsymbol{\theta_t} \mid G ^{(n)}&\widesim{iid} G ^{(n)}, \qquad  t=1,2,..,T\\
	G ^{(n)}\mid \upsilon,\sigma^2,\rho &\sim DP(\upsilon G_0^{(n)})\\
	G_0^{(n)}(\cdot \mid \sigma^2,\rho)&\equiv MVN(\cdot \mid \boldsymbol{0},\sigma^2(\boldsymbol{D}-\rho\boldsymbol{W})^{-1})\\
	\boldsymbol{\beta}&\sim MVN\left(\boldsymbol{\beta}\mid \boldsymbol{\beta_0},\boldsymbol{\Sigma_\beta} \right)\\
	\tau^2&\sim  IG(a_{\tau}, b_{\tau})\\
	\sigma^2&\sim IG(a_{\sigma}, b_{\sigma})\\
	\upsilon&\sim Gamma(a_{\upsilon}, b_{\upsilon})\\
	\rho &\sim Uniform(0,1).
	\end{align*}
	
	\begin{figure}
		\centering
		\centerline{\includegraphics[width=\textwidth,height=0.6\textheight,keepaspectratio]{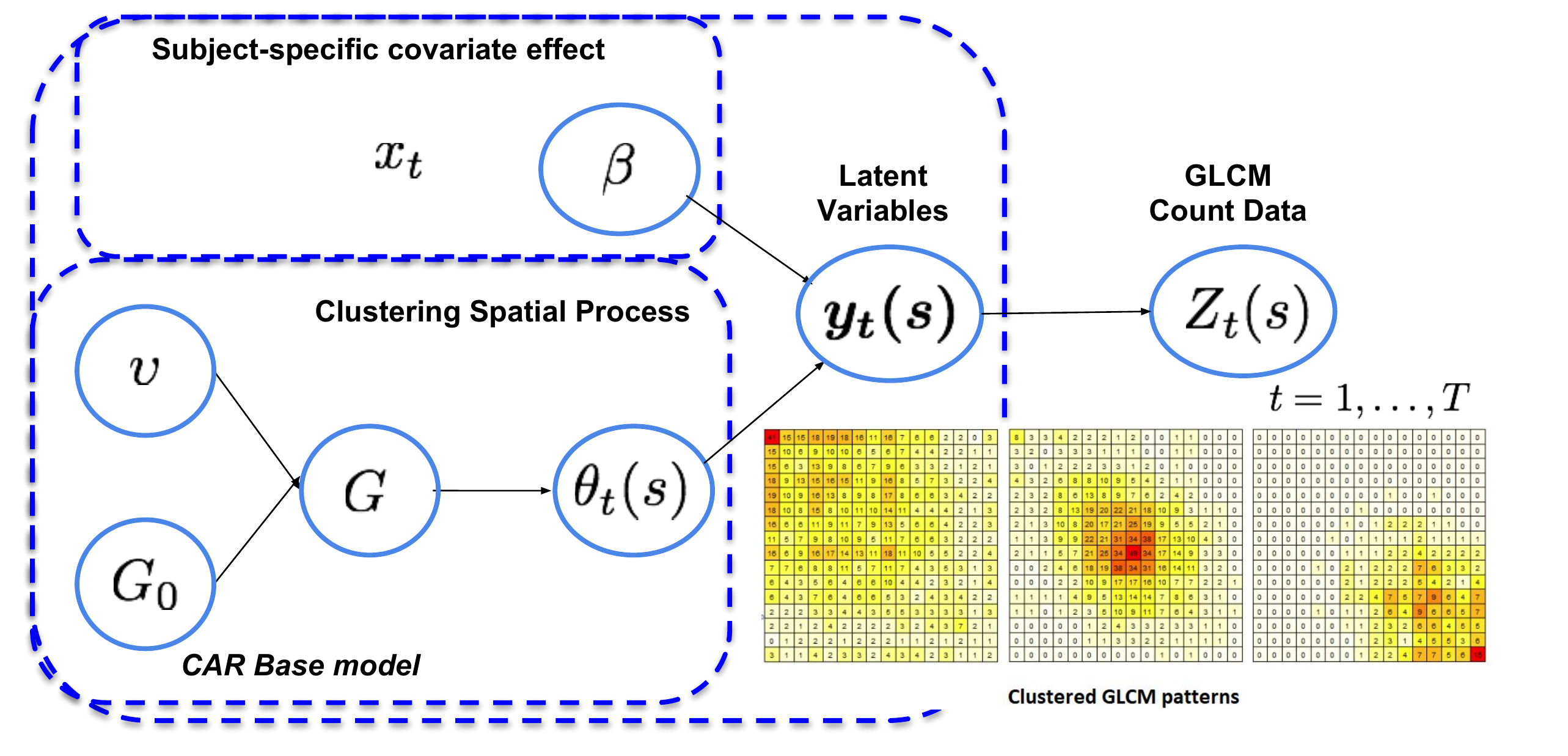}}
		\caption{Graphical representation of the proposed hierarchical rounded Gaussian SDP model.} 
		\label{fig:model}
	\end{figure}
	
\subsection{Posterior inference}
	
In this section, we outline the main steps of the MCMC algorithm used to obtain posterior samples from the proposed hierarchical rounded Gaussian SDP model, and we discuss how we can summarize the posterior inference about the heterogeneity of the GLCM objects across subjects. 
	
\subsubsection{MCMC algorithm}
The proposed MCMC sampling algorithm requires iterating the following steps:
\begin{enumerate}
\item[\textit{(a)}] \textit{Update of the latent vector $\boldsymbol{y_t}$}:
similarly as in \citet{canale2011bayesian}, we use a data augmentation step to update each $\boldsymbol{y_t}$ from a truncated Gaussian distribution, conditional on the other parameters and the cluster assignments. More specifically, 
\begin{enumerate}
\item[\textit{(a.1)}] Let $w_t$ denote the cluster assignment of subject $t$. Then,  first sample from uniform random variables
	$$u_t(s_i) \sim U \left( \Phi(a_{z_t(s_i)-1} \big| \tilde{\mu}_t(s_i),\tilde{\sigma}^2_t(s_i)),\Phi(a_{z_t(s_i)}\big| \tilde{\mu}_t(s_i),\tilde{\sigma}^2_t(s_i))\right) 
	$$
	where 
	\begin{eqnarray*}
		&&\tilde{\mu}_t(s_i) = \mu_{w_t}(s_i)+\Sigma_{w_t,12}\Sigma_{w_t,22}^{-1}\left(\boldsymbol{y_t(-s_i)^{old}}-\boldsymbol{\mu}_{w_t}\boldsymbol{(-s_i)} \right),\\
		&&\tilde{\sigma}^2_t(s_i) = \tilde{\sigma}^2_{w_t}(s_i)-\Sigma_{w_t,12}\Sigma_{w_t,22}^{-1}\Sigma_{w_t,21}
	\end{eqnarray*}
	 are the usual conditional expectation and conditional variance of the  multivariate normal with $\boldsymbol{\mu_{w_t}} = \boldsymbol{x_t}\boldsymbol{\beta}+\gamma_t\boldsymbol{\theta_{w_t}^{\ast}}$ and $\boldsymbol{\Sigma_{w_t}} = \tau^2\boldsymbol{I}$, and $\boldsymbol{-s_i}$ denotes the collection of all locations with the exclusion of the currently considered location $s_i$
	\item[\textit{(a.2)}] Based on the generated $u_t(s_i)$, update the latent $y_t(s_i)$ as follows: 
	$$y_t(s_i)^{new} = \Phi^{-1}\left(u_t(s_i) \big| \tilde{\mu}_t(s_i),\tilde{\sigma}^2_t(s_i)\right) $$
	\end{enumerate}
	
	\item[\textit{(b)}]\textit{Update of the Dirichlet process parameters $\boldsymbol{\theta_t}$ and corresponding cluster allocations $w_t$}:  we jointly update the allocation vector $w_t$ and the subject-specific parameter vectors $\boldsymbol{\theta}_t$'s as follows:
	\begin{enumerate}
	\item[\textit{(b.1)}]  Integrate out the random probability measure $G$ and use the P\'olya Urn representation of the Dirichlet Process:  
	\begin{align*}
	f\left(\boldsymbol{\theta_t} \big| \boldsymbol{\theta_{t^{'}}}, t \neq t^{'}\right) &\propto \frac{1}{q_0+\sum_{j=1}^{T_{(-t)}^{\ast}}T_{j(-t)}q_j} \big\{ q_0h(\boldsymbol{\theta_t}\mid \boldsymbol{y_t},\boldsymbol{\beta},\tau^2,\sigma^2,\rho)\\
	&+\sum_{j=1}^{T_{(-t)}^{\ast} }T_{j(-t)}q_j\delta_{\boldsymbol{\theta_j^{\ast}}}(\boldsymbol{\theta_t}) \big\} ,
	\end{align*}
	for $t=1,2,...,T$, where the subscript $\left( - t\right)$ is used to denote all relevant quantities when $\boldsymbol{\theta_t}$ is removed from $\boldsymbol{\theta}$. More specifically, $T_{(-t)}^{\ast}$ refers to the number of clusters in $\left(\boldsymbol{\theta_{t^{'}}},t^{'} \neq t \right) $, and $T_{j(-t)}$ refers to the number of elements in cluster $j$, $j = 1,...,T_{(-t)}^{\ast}$, with $\boldsymbol{\theta_t}$ removed. The term $q_j$, $j=1, \ldots, T_{(-t)}^{\ast}$  denotes the density of the multivariate Gaussian distribution with mean vector $\boldsymbol{x_t\beta}+\gamma_j\boldsymbol{\theta_j^{\ast}}$ and covariance matrix $\tau^2\boldsymbol{I}.$ Let $\boldsymbol{H}=\left( \boldsymbol{D}-\rho \boldsymbol{W} \right) ^{-1}$. Then,
	\begin{align*}
	q_0	&= \upsilon\left|(\tau^{-2}\gamma_t^2\boldsymbol{I}+\sigma^{-2}\boldsymbol{H}^{-1})^{-1} \right|^{\frac{1}{2}} \left|\boldsymbol{H} \right|^{-\frac{1}{2}}(2\pi\sigma^2\tau^2)^{-\frac{n}{2}}\\
	&\exp\left[-\frac{1}{2}\tau^{-2}(\boldsymbol{y_t-x_t\beta})^{'}(\boldsymbol{I}-\tau^{-2}\gamma_t^2(\tau^{-2}\gamma_t^2\boldsymbol{I}+\sigma^{-2}\boldsymbol{H}^{-1})^{-1})(\boldsymbol{y_t-x_t\beta})\right] 
	\end{align*}
	Similarly, $h(\boldsymbol{\theta_t}\mid \boldsymbol{y_t},\boldsymbol{\beta},\tau^2,\sigma^2,\rho)$ denotes the density of multivariate normal distribution with mean vector $\tau^{-2}\gamma_t(\tau^{-2}\gamma_t^2\boldsymbol{I}+\sigma^{-2}\boldsymbol{H}^{-1})^{-1}(\boldsymbol{y_t-x_t\beta})$ and covariance matrix $(\tau^{-2}\gamma_t^2\boldsymbol{I}+\sigma^{-2}\boldsymbol{H}^{-1})^{-1}$. Note that once $\boldsymbol{\theta_t}$ is updated, $w_t$ is also implicitly updated. 
	\item[\textit{(b.2)}] In order to improve the mixing properties of the chain, we resample the unique cluster atoms $\theta_j^*$ based on the specific configuration  $\boldsymbol{w}=(w_1,w_2,...,w_T)$ obtained in step 2(a) and the associated $T^{\ast}$ cluster values, by drawing updated cluster atoms from the full conditional $f(\boldsymbol{\theta_j^{\ast}} \big| T^{\ast},\boldsymbol{w},\boldsymbol{y},\boldsymbol{\beta},\sigma^2,\tau^2,\rho )$, which  is a multivariate normal distribution with mean vector $$\tau^{-2}(\sum_{t:w_t=j}\tau^{-2}\gamma_t^2\boldsymbol{I}+\sigma^{-2}\boldsymbol{H}^{-1})^{-1}(\sum_{t:w_t=j}\gamma_t(\boldsymbol{y_t-x_t\beta}))$$ and covariance matrix
	$(\sum_{t:w_t=j}\tau^{-2}\gamma_t^2\boldsymbol{I}+\sigma^{-2}\boldsymbol{H}^{-1})^{-1}$.
	\end{enumerate}
	\item[\textit{(c)}] \textit{Update the vector of regression coefficients $\boldsymbol{\beta}$.} \\The full conditional for  $\boldsymbol{\beta}$  is a multivariate normal distribution  with mean\\ $\left(\boldsymbol{\Sigma_\beta^{-1}}+\tau^{-2}\sum_{t=1}^{T}\boldsymbol{x_t^{'}x_t}\right)^{-1}
	\left(\boldsymbol{\Sigma_\beta^{-1}}\boldsymbol{\beta_0}+\tau^{-2}\sum_{t=1}^{T}\boldsymbol{x_t^{'}}(\boldsymbol{y_t}-\gamma_t\boldsymbol{\theta_t})\right)$\\ and variance $\left(\boldsymbol{\Sigma_\beta^{-1}}+\tau^{-2}\sum_{t=1}^{T}\boldsymbol{x_t^{'}x_t}\right)^{-1}$.\\
	
	\item[\textit{(d)}] \textit{Update the sampling variance $\tau^2$.}
	The full conditional distribution for $\tau^2$ yield an inverse-gamma distribution $IG(\tau^2|\tilde{a}_{\tau},\tilde{b}_{\tau})$ with $\tilde{a}_{\tau}=\frac{Tn}{2}+a_{\tau},
	\tilde{b}_{\tau}=b_\tau+\frac{1}{2}\sum_{t=1}^{T}(\boldsymbol{y_t-x_t\beta}-\gamma_t\boldsymbol{\theta_t})^{'}(\boldsymbol{y_t-x_t\beta}-\gamma_t\boldsymbol{\theta_t})$\\
	\item[\textit{(e)}]\textit{Update the CAR variance $\sigma^2$}. Given that marginalize $G$, the $\boldsymbol{\theta_j^{\ast}}$ are independent and identical with distribution $G_0$, the full conditional distribution for $\sigma^2$ is given by an inverse-gamma distribution $IG(\sigma^2|\tilde{a}_{\sigma},\tilde{b}_{\sigma})$ with $\tilde{a}_{\sigma}=\frac{T^{\ast}n}{2}+a_{\sigma},$ and $\tilde{b}_{\sigma}=b_\sigma+\frac{1}{2}\sum_{j=1}^{T^{\ast}}(\boldsymbol{\theta_j^{\ast}})^{'}\boldsymbol{H}^{-1}(\boldsymbol{\theta_j^{\ast}})$\\

	\item[\textit{(f)}] \textit{Update the spatial coefficient in the CAR precision matrix, $\rho$.} Since $\rho$ appears in the off-diagonal part of $H$, Adaptive Rejection Metropolis Sampling method \citep{gilks1995adaptive},  developed from Adaptive Rejection Sampling \citep{gilks1992adaptive} by including a Metropolis step to accommodate non-concavity in the log density, is implemented to draw posterior samples from the target density 
	$\left|H \right| ^{-\frac{T^{\ast}}{2}} \exp\left(-\frac{\sigma^{-2}}{2}\sum_{j=1}^{T^{\ast}}(\boldsymbol{\theta_j^{\ast}})^{'}\boldsymbol{H}^{-1}(\boldsymbol{\theta_j^{\ast}}) \right)$\\

	\item[\textit{(g)}] \textit{Update the DP concentration parameter, $\upsilon$.}
	In order to simulate posterior samples for $\upsilon$, we follow the augmentation idea from Escobar and West (1995): we introduce an auxiliary variable $\eta$ sampled  from a beta distribution $(\eta| \upsilon^{old}+1, T)$, then  $\upsilon$ is updated from the two-component mixture of gamma distributions $p\cdot Gamma(a_\upsilon+T^{\ast},b_\upsilon-log(\eta) )+(1-p)\cdot Gamma(a_\upsilon+T^{\ast}-1,b_\upsilon-log(\eta))$, where $p=( a_\upsilon+T^{\ast}-1) \Big/ \Big( T(b_\upsilon-log(\eta))+a_\upsilon+T^{\ast}-1\Big).$
\end{enumerate}
	
\subsubsection{Cluster inference}\label{sec:clusterinference}
Characterization of intra-lesion pattern heterogeneity is the predominate motivation for study of GLCM objects. 
The use of a spatial Dirichlet process prior provides a flexible method  for unsupervised learning with GLCM object data.  Several approaches have been proposed in the  literature to conduct post-MCMC inference with Bayesian nonparametrics. This often requires addressing either the issue of label switching intrinsic to Bayesian mixture models \citep{jasra2005markov} and/or the clustering behavior of the Dirichlet Process \citep{miller2017mixture}.
For example, \citet{medvedovic2002bayesian} employ a classical agglomerative hierarchical clustering method on the basis of dissimilarity matrices, with elements calculated as the posterior probability that any two observations $i$ and $j$ are assigned to different mixture components. 
Also \citet{bandyopadhyay2016non} employ hierarchical clustering  to cluster ordered periodontal measurements within and across subjects  on the basis of a dissimilarity matrix. \citet{dahl2006model} and \citet{dahl2007multiple} estimate cluster membership by minimizing the posterior expected loss of binding through processing of all pairwise posterior probabilities from the proportion of MCMC samples in which two subjects are assigned the same mixture component. Whenever the optimal partition is not constrained to be in the sample, \citet{hastie2015sampling} have found that leveraging the entire sample of MCMC draws though implementation of additional clustering methods (such as partitioning around medoids) is more favorable in reducing 
Monte Carlo error.  \XL{More recently, \citet{wade2018} and \citet{RastelliFriel2018} have proposed decision theoretic approaches to identify optimal partitions.  \citet{wade2018} also define posterior credible balls to quantify the uncertainty of the estimated partitions. Here,  we propose an approach similar as in \citet{zhang2016spatiotemporal}, who applied a hierarchical clustering approach to fMRI studies.
The approach maintains simplicity to facilitate communication and discussion of the results with medical doctors, while focusing on the posterior estimates of the latent GLCM patterns to organize the observed GLCM matrices into homogeneous groups.  More specifically, we first} obtain subject-level posterior mean surfaces of the random effects \XL{using} all iterations after burn-in.  We then compute  a dissimilarity matrix using squared Euclidean distance between each pair of the subjects. Finally, we obtain a partition of the GLCM matrices by using the \citet{ward1963hierarchical} clustering method with Ward's minimum variance implemented recursively by the Lance-Williams algorithm \citep{murtagh2014ward} and \citet{krzanowski1988criterion} criteria which minimizes intra-cluster dispersion. 

\section{Case study: textural patterns in adrenal lesions}
\label{case_study}
In this section, we apply our proposed hierarchical rounded Gaussian SDP method and present the analysis of the retrospective adrenal diagnostic study data. The objective is to identify intrinsic textural patterns observable with CT as well as characterize their heterogeneity within the studied cohort. To assess whether the identified patterns are pathologically oriented, we study the relationships between GLCM lesion subtypes and pathology endpoints.
	
In particular, the proposed hierarchical rounded Gaussian SDP method was implemented to obtain a GLCM lesion subtype for each lesion for both NC and DL phase scans without consideration of true pathology status. 
We considered the (unnormalized)  sum of counts of the GLCMs objects as a subject-specific covariate $\mathbf{x}_t$ in the mixed effects model for the latent $\boldsymbol{y(s)}$, to acknowledge that larger GLCMs are characterized by larger counts. We further assumed a vague prior for $\boldsymbol{\beta}$, by setting \XL {$\boldsymbol{\Sigma_0}= 10^5\boldsymbol{I}$}. The hyperparameters of the priors on $\tau^2$ and $\sigma^2$ were set to $a_{\tau} = b_{\tau} = a_{\sigma} = b_{\sigma} = 0.0001$. A standard choice of a $Gamma(a_\upsilon,b_\upsilon)$ prior with $a_\upsilon=b_\upsilon=1$ is imposed for $\upsilon$, although the results appeared robust to fixing $\upsilon=1$ \citep{quintana2016bayesian}. For each simulation, we ran the MCMC with 20,000 iterations, discarding the first 10,000 iterations as burn-in. Convergence was conferred for each replicate study sample from the Geweke diagnostic \citep{citeulike:1176289} as implemented in the R package `coda'.
	
	\begin{figure}
		\centering
		\includegraphics[width=\textwidth,height=0.95\textheight,keepaspectratio]{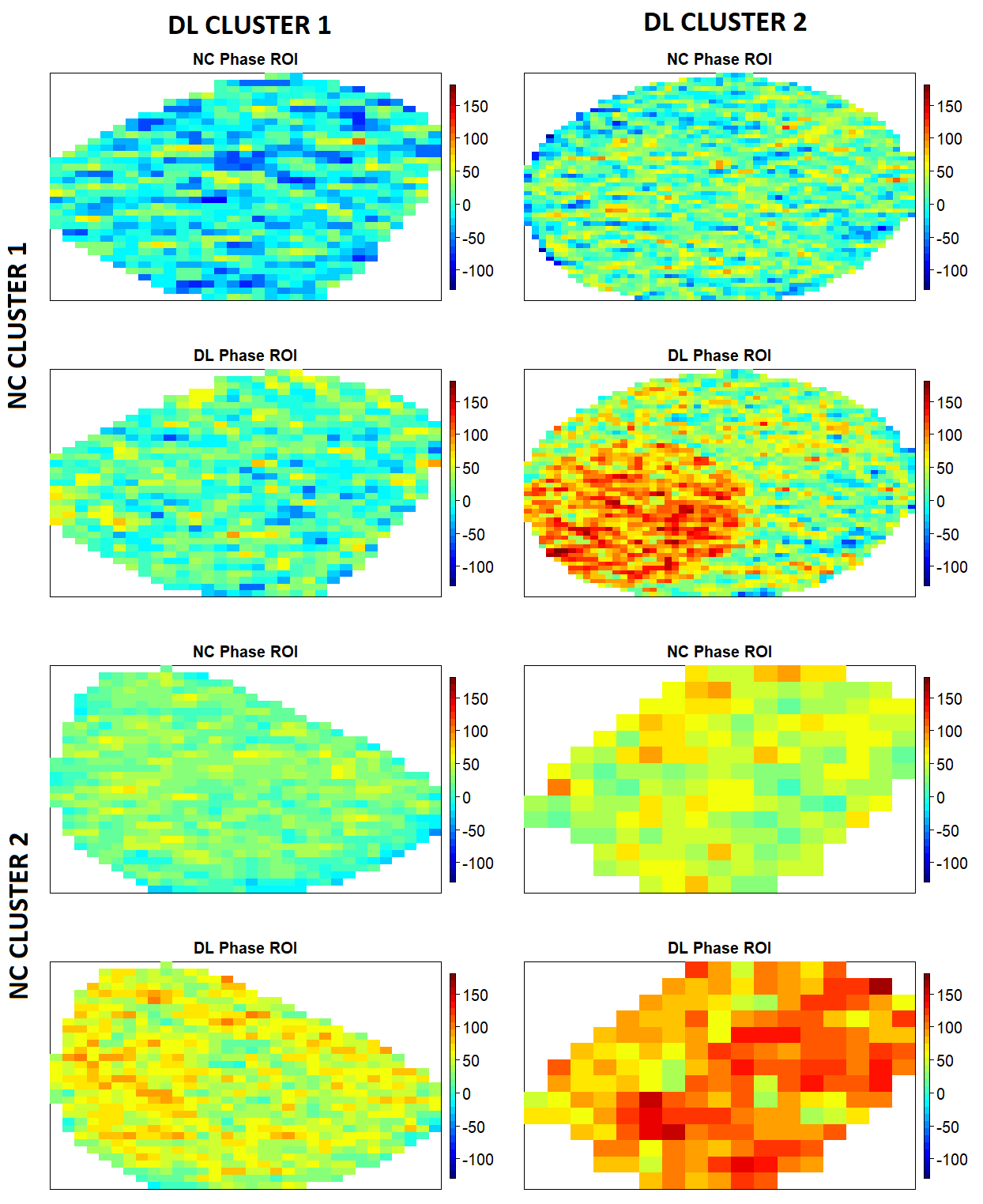}
		\caption{\small Pixel-level ROIs of representative subjects for the 4 resultant subgroups, where NC stands for noncontrast scan phase, DL stands for delay scan phase.}
		\label{fig:case-study_1}
	\end{figure}
    
    \begin{figure}
		\centering
		\includegraphics[width=\textwidth,height=0.94\textheight,keepaspectratio]{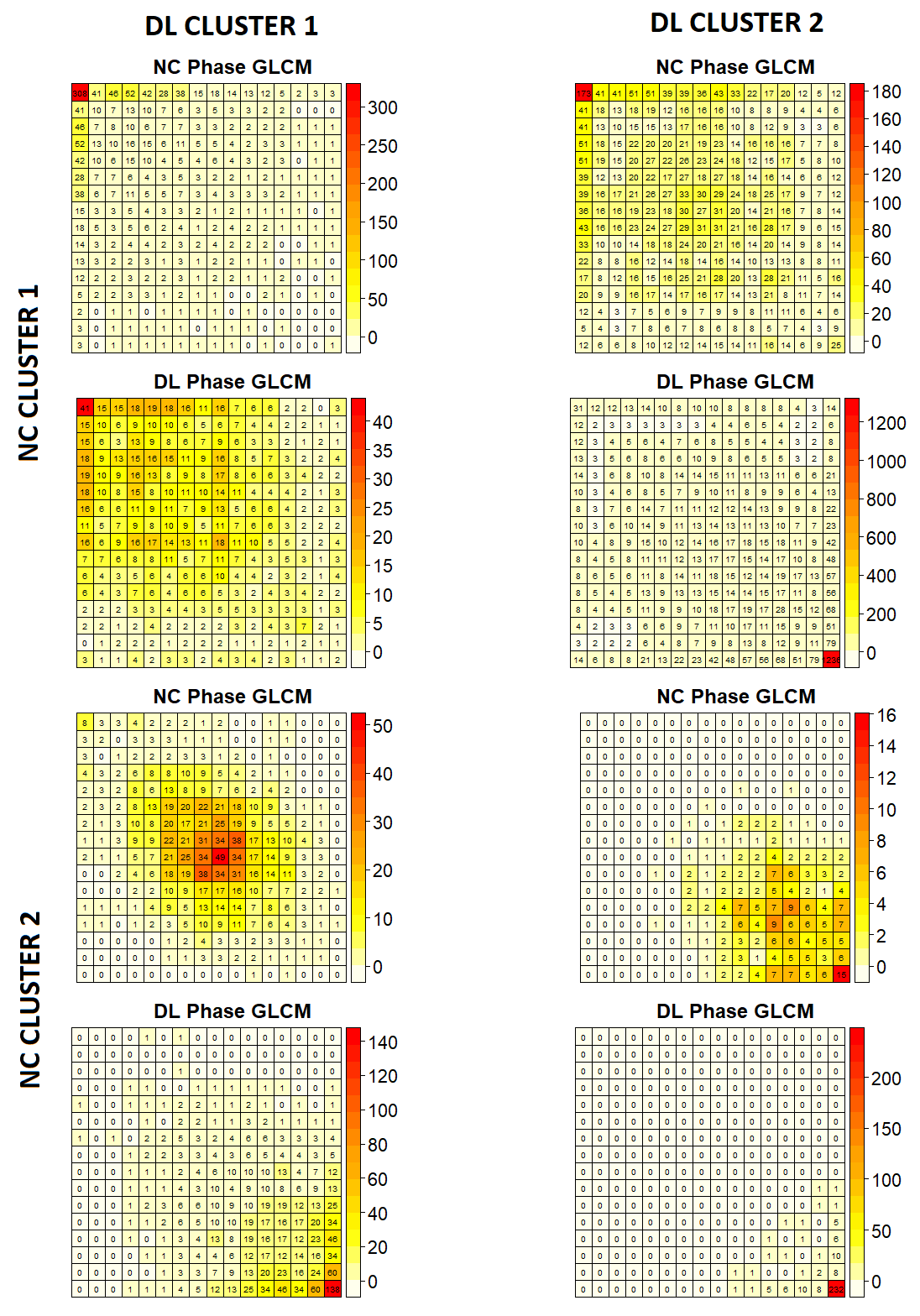}
		\caption{\small GLCMs of representative subjects for the 4 resultant subgroups, where NC stands for noncontrast scan phase, DL stands for delay scan phase. Each GLCM corresponds to each ROI in Figure~\ref{fig:case-study_1}. } 
        \label{fig:case-study_2}		
	\end{figure}
	
	The MCMC-based inference described in Section 4.2 identified two clusters of patients for each type of CT scan (NC and DL). Figure~\ref{fig:case-study_1} and Figure~\ref{fig:case-study_2} depict scans and GLCMs for representative lesions observed within the resultant four subgroups, respectively. The subgroup defined by NC cluster 1 and DL cluster 1 (top row, upper left quadrant), is characterized by high NC peak cells in the north-west direction and high DL peak cells either in the north-west direction or center, which illustrates that the majority of co-occurrences map to low grey-levels. 
	By way of contrast, the subgroup defined by NC cluster 2 and DL cluster 2 (lower right quadrant) has both high peak cells in the south-east direction, demonstrating prevalence and co-occurrence of high grey-levels, as higher HU intensity are observed in the original NC and DL lesion ROI images.
	In addition, the DL GLCMs tend to be very sparse, yielding many 0s within low grey-level pairs. The subgroup defined by NC cluster 1 and DL cluster 2 has peak counts for NC in the north-west direction and south-east direction for DL, representing lower HU intensity in NC lesion ROI and higher HU intensity in DL lesion ROI. Finally, the subgroup  defined by NC cluster 2 and DL cluster 1 is characterized by peak counts for NC in the central and high DL peak counts in the south-east cells, elucidating median HU intensity for NC scans and high HU intensity for DL scans. Furthermore, we should note that the sparsity in the DL GLCM measurements observed in the bottom row GLCMs is not revealed in the other subtypes.
	
Table~\ref{mytable} reports the resultant combined cluster assignments for combined non-contrast and delay images, together with the number of true malignant lesions in that specific subgroup. While entirely unsupervised, e.g. fitted to the GLCM matrices without reference to pathology, the table suggests that patterns identified by the Bayesian nonparametric model may convey pathological  orientation. 
Flagging lesions as malignant (benign) when assigned to subtypes where the fraction of true malignancies is more (less) than $50\%$, only $21.9\%$ of lesions are assigned to discordant subtypes. 

Additional findings for individual scans (NC and DL) are presented in Section A of the supplementary materials.	
	%

    \begin{table}
		
		\caption{\label{mytable}{\bf Adrenal case study results:} Cluster assignments for combined non-contrast and delay images. The number in the brackets in each cell represents the number of true malignant lesions in that specific subgroup.}
		\centering
		
		\begin{tabular}{llcc}
			&           & \multicolumn{2}{c}{Non-contrast scan}                                        \\
			&           & \multicolumn{1}{l}{cluster 1} & \multicolumn{1}{l}{cluster 2} \\
			\multicolumn{1}{c}{\multirow{2}{*}{Delay scan}} & cluster 1 & 59(2)                         & 52(20)                        \\
			\multicolumn{1}{c}{}                    & cluster 2 & 5(2)                          & 94(72)                       
		\end{tabular}
	\end{table}

\section{Simulation study}
\label{Sec:simulation}
This section describes a simulation study devised to evaluate the performance of the proposed hierarchical rounded Gaussian spatial Dirichlet process (HRGSDP) method with respect to several competing clustering approaches typically encountered in the analysis of imaging data in oncology studies.
	
	\subsection{Simulation Design}
	To obtain appropriate synthetic data capable of representing the spatially-oriented patterns within GLCMs exhibited in our diagnostic cancer study of adrenal lesions, we generated GLCMs constructed with 16 gray-levels with normalized element-wise probability densities. Our simulation study was conducted using two data generating models. The first model assumed that density patterns within the GLCM objects were a sample from a bivariate normal distribution with \XL{ 
	$$\mu=(2+c,14-c)^{'} \quad \text{and} \quad \Sigma
	=s 
	\begin{bmatrix}
	1 & -0.7 \\
	-0.7 & 1\\	
	\end{bmatrix}.$$}
	Various simulation scenarios were obtained as different combinations of the shift parameter $c$ and the scale parameter $s$, as described further below. A final smoothed event rate surface, wherein the event rate is regarded as the probability of observing a co-occurrence count within a given cell, was obtained for each simulated GLCM by smoothing over the Gaussian-derived empirical rate surface calculated in proportion to the number of generated points in each grid and with respect to the total number of generated points of the entire grid surface. Additionally, to reflect inter-patient heterogeneity in image size we scaled the rate surface over the discrete GLCM space by a random integer that is sampled from 500 to 20000. The scale densities were then rounded to the nearest integer value to yield simulated counts. The simulation design yields a distribution over the GLCM space that generates reasonably small and large GLCM count surfaces, representing both over- and under-dispersed scenarios. 
	Simulation scenarios were then obtained by generating GLCMs under different choices of the mean shift parameter $c$ and variance scale parameter $s$. In particular, for a given $s$ we generated GLCMs as a mixture of five components represented by $c \in \{5, 5.5, 6, 6.5, 7\}$, such that the center of the generating bivariate normal distribution shifts their location gradually from north-west toward south-east along the $135^\circ$ diagonal of the GLCM. We also evaluated performance by varying {\XL{$s \in \{10, 12, 15, 16, 17\}$}}, with each value of $s$ representing the extent to which the noise covers the true signal in the underlying spatial patterns intrinsic to each tissue class. As we set the mean vectors of the bivariate normal distribution of the different components closer to each other, increasing the value of $s$ diminishes the extent to which the spatial patterns are differentiable by any method. For each $s$, we generated 100 subjects in total, with 20 patients in each of the five subtypes defined by $c$. {\XL{ To interrogate robustness to the Gaussian assumption, in the Supplementary material, Section B, we also present a simulation using the bivariate skew normal distribution of \cite{azzalini1999statistical} as the latent generator. The results are consistent with those described below for the bivariate normal setting.}}

	\subsection{Comparison with competing methods for texture analysis}
    \label{comp-meth}
    We fitted the simulated GLCMs using the proposed HRGSDP method, where the model hyper-parameters were set to obtain vague prior specification as in the case study. MCMC was implemented with 20,000 iterations, the first 10,000 of which were discarded as burn-in. Convergence was assessed by  using the Geweke diagnostic [31] with implementation using the R package `coda'. 
	
	We then considered five alternative clustering approaches, which are typically employed for the analysis of imaging data in oncology studies.
	More specifically, we considered (i) hierarchical clustering based on a few GLCM-derived texture features, which are reported routinely in radiomics studies (such as contrast, correlation, homogeneity, energy and entropy) \citep{kumar2012radiomics}. We refer this method as `FeaHC'. Additionally, for the same GLCM-derived texture features,  (ii) spectral clustering, referenced as `FeaSC', (iii) K-means clustering, referenced as `FeaKM' (iv) consensus clustering algorithms, referenced as `FeaCO' and (v) Gaussian mixture model, referenced as `FeaGMM' were also implemented as typically reported in clinical imaging studies \citep{gensheimer2015assessing,wu2016robust,parmar2015radiomic,zhou2017radiomics}.  
	
	In contrast to our unsupervised approach, all of the competing methods above require the pre-specification of the number of clusters in advance of analysis. In order to optimize their performance , we adopted the true number of clusters as the ground truth, i.e. we assumed five simulated subtypes. That is, clusters were obtained for each simulated dataset by cutting the resultant dendrogram with rank equal to the true number of clusters. Of course, in real data settings, such knowledge would not be available. By way of contrast, no truth-based specification was used for ascertaining subtypes using our Bayesian method, which provided an artificial advantage favoring competing approaches.
	
	\subsection{Performance evaluation} 
	While our modeling framework was devised for the purpose of unsupervised clustering, comparing the relative quality of competing sets of unsupervised clusters is abstruse in the absence of an underlying mechanism for understanding realizations of the objects that are the basis for cluster analysis. Therefore, our simulation study was devised under the assumption that GLCM objects arise from disparate true classes of lesions.  Each class was assigned a true probability distribution for generating individual GLCMs, as described in Section 5.1, providing a basis for comparison. Specifically, method performance was evaluated in relation to the matching matrix constructed with true memberships as rows and the predicted membership as columns such that the $(i,j)$th entry measures the counts of the $i$th class subject with membership assigned by the clustering approach $j$. The matching matrix was computed with our proposed method and the above mentioned approaches for each simulated cohort on 100 replicate simulations for each value of $s.$ To evaluate and compare the performance of our proposed HRGSDP method and the other competing approaches mentioned above, we used two performance metrics: \XL{a measure of the} mis-assignment rate and the Pearson chi-square test statistic obtained from the resultant matching matrices. 
	
	With our proposed Bayesian nonparamteric approach, a membership assignment was obtained for each subject by posterior clustering of $\boldsymbol{\theta}$ (see Section \ref{sec:clusterinference}). \XL{Then, the assigned membership along with its true class (defined by $c$) were interrogated to measure the extent to which GLCMs arising from truly identical pattern distributions tended to cluster together. The competing methods require the pre-specification of the total number of clusters to assign the GLCMs to cluster labels. However, the HRGSDP is unsupervised, as it provides a data-driven estimate of the cluster rank. Nevertheless, in the following, we measure the performance of the competing methods with respect to the true rank (5) used to simulate GLCM patterns, which would be unknown in practice.}
	
	\XL{Methods were compared on the basis of their mis-assignment rate, which was defined to characterize the proportion of GLCM objects that cluster with truly uncommon patterns.  This is calculated by tabulating the distribution of model-derived predicted cluster assignments by their true class, and identifying the mode class for each cluster (class with maximum proportion). For each predicted cluster, the number of assignments to other classes (not the mode) is taken as an error. The errors are summed and normalized by the sample size.
	Then, the resultant mis-assignment rate was defined as the proportion of mis-assigned subjects overall.}
	
	The Pearson chi-square test statistic was also used to measure the homogeneity of true class membership and the assigned class membership. 
	Thus, larger values of the test statistic, which represent larger deviations from the null random frequency distribution, characterize better performance as cluster assignments derived from clustering techniques become more consistent with their true subtype memberships.
	
	\subsection{Simulation results}
	{\XL{Figure \ref{fig:simu1} reports the  mis-assignment rate distributions obtained in our simulation study as a function of the  values of the dispersion parameter 
	$s \in \{10,12,15,16,17\}$. 
	\begin{figure}
		\centering
		\centerline{\includegraphics[width=1.05\textwidth,height=\textheight,keepaspectratio,angle=-90]{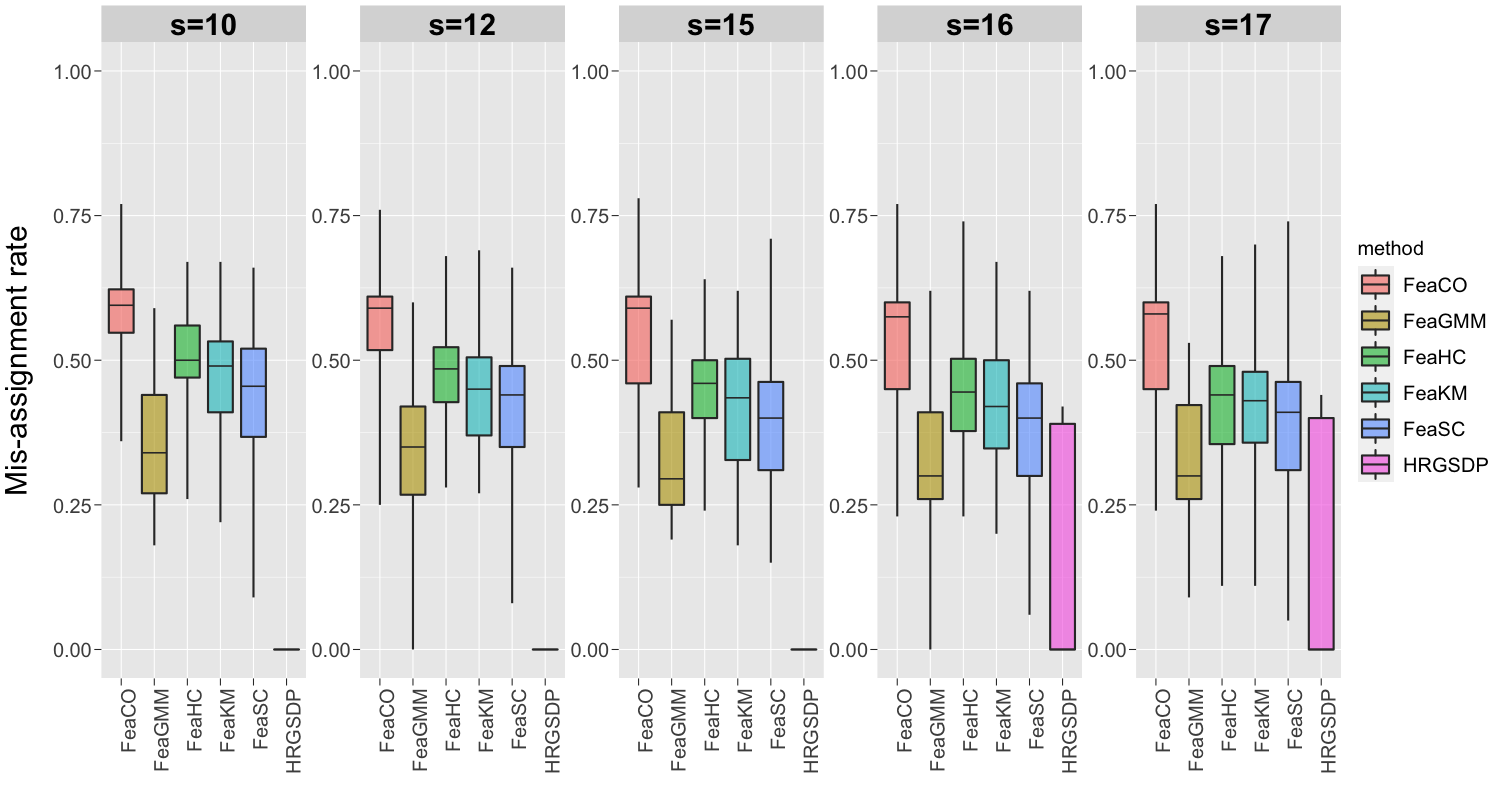}}
		\caption{{\bf Simulation results.} Mis-assignment rate obtained for all methods considered in the simulation study, as a function of the sparsity parameter $s \in  \{10, 12, 15, 16, 17\}$. See section~\ref{comp-meth} for detailed explanation of the method acronyms.}\label{fig:simu1}
	\end{figure}
	Among all currently employed feature-based approaches, the mean mis-assignment rates are similar, ranging from $32.4\%$ to $58.7\%$, with the Gaussian mixture model (FeaGMM) having the best performance with $s=15$ ($32.4\%$). The best performance for consensus clustering (FeaCO), hierarchical clustering (FeaHC), K-means (FeaKM) and spectral clustering (FeaSC) methods are $53.9\%$ with $s=17$, $42.8\%$ with $s=17$, $41.9\%$ with $s=16$ and $38.6\%$ with $s=16$, respectively.  
	In all simulated scenarios, the Gaussian mixture model (FeaGMM) method yields the minimum mis-assignment rate among the feature-based approaches. 
	}}
{\XL{As a contrast, the performance of the proposed method is globally superior at smaller values of $s$ (0.1\% for $s=10$ and $5.8\%$ for $s=15$) ,
i.e. for high signal-to-noise scenarios. As the random noise increases, the performance of the proposed method decreases  but it is still better than or at least comparable to the performances of the competing methods, resulting in mean mis-assignment rates of \XL{$12.5\%$ for $s=16$ and $19.2\%$ for $s=17$.}
This behavior appears to be due to the combination of two factors. On the one hand, high noise results in  imprecise cluster estimates. On the other hand, for large values of $s$, the noise dominates the signal conveyed by the true mean spatial patterns of dependencies that characterize the  true GLCM subtypes. As our simulations demonstrate, different subtypes are 
	{\XL {best}} identified by large differences in the values of the mean location parameter $c$.
}}
	
	Figure \ref{fig:simu2} reports the distributions of the Pearson chi-square test statistic obtained as a function of the dispersion parameter {\XL{$s \in \{10,12,15,16,17\}$}}, respectively. 
	\begin{figure}
		\centering
		\centerline{\includegraphics[width=1.05\textwidth,height=\textheight,keepaspectratio,angle=-90]{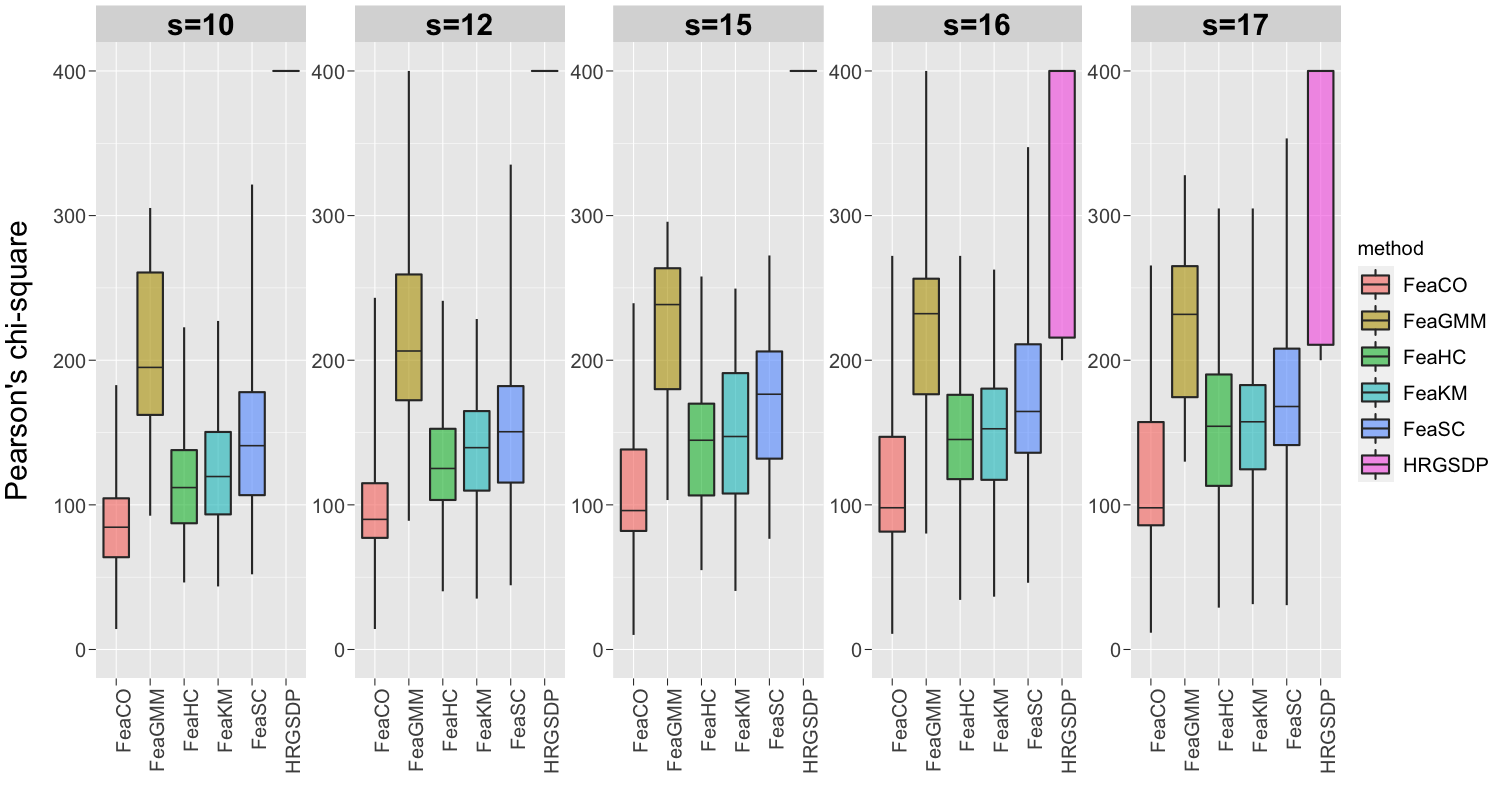}}
		\caption{{\bf Simulation results.} Pearson chi-square test statistic obtained for all methods considered in the simulation study as a function of the sparsity parameter $s \in  \{10, 12, 15, 16, 17\}$. See section~\ref{comp-meth} for detailed explanation of the method acronyms.} \label{fig:simu2}	
	\end{figure}
	{\XL{We should note that, in our simulation scenario, the maximum possible value of the Pearson chi-square test statistic is 400. This is obtained whenever all clusters identified by a model consist entirely of GLCMs arising from the same true pattern distribution. We refer to this as perfect discrimination. The means of the statistic for all feature-based approaches are similar, ranging between 83 and 225. Among them, the  Gaussian mixture model yielded the best clustering performance, reaching a maximum of 225 with $s=17$. The maximum values of the Pearson chi-square test statistic obtained for the consensus clustering (FeaCO), hierarchical clustering (FeaHC), K-means (FeaKM) and spectral clustering (FeaSC) methods were 124, 162, 160 and 177 at $s=17$, respectively.}} 
	
	Results for the HRGSDP model far exceeded those resulting from the above machine learning clustering approaches at all levels of signal-to-noise. As a matter of fact, results were perfect for $s<15$ with our proposed HRGSDP model resulting in a mean Pearson chi-square test statistic of 400 at $s=10$ and $s=12$. {\XL{In the presence of higher levels of random noise, the statistic was 373 at $s=15$, 341 at $s=16$ and 309 at $s=17$, and thus globally superior.}} 
	
	The trends in the diagnostic performance that the HRGSDP method exhibited in our simulation study conform to intuition as it pertains to identification of spatial discriminatory patterns and random noise. Overall, the proposed HRGSDP method outperformed the machine-learning methods currently used in the literature for reasonable signal-to-noise ratios. The algorithmic-oriented techniques fail to take into account the multivariate lattice structure intrinsic to GLCMs, and thus are unable to properly capture the spatial dependencies and the distributional heterogeneity inherent to a sample of GLCM objects. In the presence of high signal-to-noise in the true GLCM generating distribution, as characterized by small values of $s$, the patterns conveyed in GLCMs were more separable from each other. This resulted in enhanced discrimination results for the HRGSDP method which leverages the spatial location information. As the variance scale parameter $s$ increases, the distinction of spatial patterns among different $c$ are diminished. 
	\XL{The methods yielded similar results when adopting the skew bivariate normal distribution as the latent generator. 
	Full results arising from the skew normal distribution are described in detail in Section B of the Supplementary materials. }

\section{Discussion} 
	
	Recent emphasis with respect to improving characterizations of tumor phenotypes based on scanning technologies has produced numerous machine learning subtyping schemes that utilize summary statistics obtained from gray-level co-occurrence matrices. Two limitations confront these methods. First, they don't admit a model based framework for unsupervised learning. Moreover, they fail to fully utilize the spatial arrangements present in the image/GLCM.

	In this manuscript, we considered  the spatial distribution of enhancement levels within a lesion image as structured count data encoded by the GLCM object. The proposed Hierarchical Rounded Gaussian Spatial Dirichlet Process framework takes into account the spatial dependence observed in the GLCM lattices, and further captures the heterogeneity of texture patterns encountered within a study population, while adjusting for the individual lesion image sizes. By means of an application to an adrenal lesion study and to a synthetic dataset, we show how the proposed methodology leads to improved performance in discerning heterogeneities in textural patterns of adrenal lesions. Improvement was also demonstrated as it pertains to identifying discrete patterns that are intrinsic to adrenal lesion textural distributions with respect to commonly employed machine-learning methods for reasonable signal-to-noise ratios. The proposed hierarchical rounded Gaussian SDP model is also capable of dealing with \XL{ the skewness and zero-inflation} 
	of counts observed within a GLCM, offering flexibility to deal with correlated multivariate count data. A salient characteristic of the proposed methodology is that, by considering the entire GLCM lattice, our method avoids the processing and analysis of potentially redundant feature-sets, which often require ad-hoc variable/feature selection steps for analysis that are targeted to each specific application. Therefore, we believe that our procedure may help increase the reproducibility of radiomics studies. Overall, the methodology offers insights into the manner in which imaging data may be better utilized to identify complex patterns that characterize the intrinsic heterogeneity observed in tumor pathology.
	
    Adrenal lesion diagnosis remains a challenge for radiologists. In the context of our case study, images were selected for which routine radiological diagnosis approaches a random guess. Our proposed approach, on the other hand, capable of capturing spatial patterns in patients' scans, yielded patterns from unsupervised learning, which described malignant and benign majority subtypes when combined with pathology.
    
    We should additionally note that due to sparse GLCMs observed in our data, several GLCM summary features proposed in the radiomics literature were not defined from the actual lesion scans observed in our objective case study. For instance, some GLCMs, especially the delay phase scans, are characterized by a single non-zero entry. In such a scenario, a typical GLCM derived feature such as the correlation is not obtainable. Our Bayesian nonparametric approach, however, handles sparsity in GLCM observables facilitating robustness for actual application. 
	
	A few limitations should be noted, however. The clustering performance of the proposed hierarchical rounded Gaussian SDP method depends upon the extent to which functional patterns in the GLCM lattice are separable among subtype objects. 
	As intrinsic to all Bayesian approaches, both prior and hyperprior specification is required for inference.  {\XL{The use of a CAR prior to describe global spatial patterns in the GLCMs is computationally convenient, but it may result in over-smoothing. While this was not the case in our application -- and some smoothing may be beneficial to cluster the GLCMs' patterns -- \citet{DuncanMengersen2020} show how different combinations of the parameter values $a_\sigma$ and $b_\sigma$  can be used to fit models with varying degrees of smoothing. They also introduced ``Goodness-of-smoothing" metrics to compare models.  Alternatively,    one could model the spatial dependence as in \citet{bandyopadhyay2016non}  via a G-Wishart prior incorporating the CAR specification. A further limitation of our approach is that it does not provide an assessment of the uncertainty around the estimated clusters. \cite{wade2018} have recently characterized  95\% credible balls obtained based on suitable metrics on the space of partitions. The use of  hierarchical clustering  allows for easy communication and discussion of the results with radiologists, but appropriate ways to summarize uncertainty for diagnosis purposes are needed, especially in high-dimensional settings as the ones considered here.
	Possible extensions of our approach include embedding the hierarchical spatial clustering approach with other flexible models devised for quasi-sparse count data \citep[see, e.g.][]{Datta2016}.  Finally,}} while unsupervised learning facilitates robustness to overfitting, supervised approaches may be developed 
	within the proposed Bayesian framework by calibrating the Bayes factor for each class membership \citep{fronczyk2015bayesian, wang2020functional, wang2019efficient}.
	

\bibliographystyle{rss}

\bibliography{ref}
\end{document}